\documentclass{article}

\usepackage{url}
\usepackage{amsmath}
\usepackage{amssymb}
\usepackage[pdftex]{graphicx}
\usepackage[linesnumbered,ruled,lined]{algorithm2e}
\usepackage{algpseudocode}
\usepackage{epstopdf}
\usepackage{authblk}
\usepackage{setspace}

\title{Schema Integration on Massive Data Sources}
\author[a]{Tianbao~Li}
\author[a]{Hongzhi~Wang \thanks{Corresponding author: wangzh@hit.edu.cn}}
\author[a]{Jianzhong~Li}
\author[a]{Hong~Gao}
\affil[a]{Harbin Institute of Technology, Harbin 150001, China}

\begin{document}
	\begin{spacing}{1.0}
	\maketitle
	
	\begin{abstract}
		As the fundamental phrase of collecting and analyzing data, data integration is used in many applications, such as data cleaning, bioinformatics and pattern recognition. In big data era, one of the major problems of data integration is to obtain the global schema of data sources since the global schema could be hardly derived from massive data sources directly. In this paper, we attempt to solve such schema integration problem. For different scenarios, we develop batch and incremental schema integration algorithms. We consider the representation difference of attribute names in various data sources and propose ED Join and Semantic Join algorithms to integrate attributes with different representations. Extensive experimental results demonstrate that the proposed algorithms could integrate schemas efficiently and effectively.\\
		keywords: Information integration, Schema mapping, Schema integration.
	\end{abstract}

\section{Introduction}

Nowadays, massive data sources are distributed on the Web. To make sufficient use of information in such data sources, information integration is in demand. Information integration merges information from heterogeneous sources with differing conceptual, contextual and typographical representations\footnote{\url{http://en.wikipedia.org/wiki/Information_integration}}. In database community, information integration often provides a uniform interface for heterogeneous data sources.

Considering of its importance, information integration has been widely studied and many techniques have been proposed. However, existing techniques are not suitable for the integration of massive data sources in big data era due to the absence of the global schema.

A traditional information integration system often requires a predefined global schema, and schema mapping techniques are applied to map local schemas of heterogeneous data sources to the global schema. In contrast, during information integration on massive data sources, it is difficult for users to predefine the global schema and the relationship between the global schema and each local schema, since it is difficult for users to capture the whole view of all the data sources.

Thus, schema integration, which is to generate a global schema for all data sources with the corresponding relationship between the global schema and local schemas, is an essential step for information integration on massive data sources. It brings following challenges.

On one hand, schema integration may be misled by synonyms, homonyms and the misspellings in the attribute names in different schemas. Thus, the quality of integrated schema will be affected. For example, ``capable'' and ``competent'' are very similar in semantics, but they are not similar literally. On the contrary, ``exportable'' and ``importable'' are opposite to each other, but they just look similar in spellings. To generate a high-quality schema, synonyms, homonyms and the misspellings in different schemas have to be handled. This challenge involves the identification of synonyms and homonyms as well as the approximate matching in the attribute names in schemas to achieve effective schema integration.

On the other hand, schema integration on massive data source requires handling a large set of schemas with even billions of attributes. It is a costly job to identify the attributes with the same semantics. Heavy operators such as similarity join and entity resolution are in demand. Thus, the second challenge is to accomplish schema integration on massive data set efficiently.

Facing these challenges, in this paper, we study the problem of schema integration of millions even billions attributes. We consider both efficiency and effectiveness issues.

For the effectiveness issue, we design approximate matching algorithm in schema integration. With the consideration that there is no extra knowledge, it is difficult to identify synonyms and homonyms just form the characters in words. So, we bring in the knowledge base, which contains concept relationships between different names. With the knowledge base, the semantic similarity between two attributes could be evaluated, and thus semantic relationship could be identified.

For the efficiency issue, we develop efficient algorithms, adapting traditional join operation to our problem. For a set of schemas, join operation in our algorithm is used to aggregate background knowledge in avoidance of simply scanning. Also, as the large-sized data have to be stored on disks, it is crucial to decrease the time of accessing the disk. Hence, we cluster the related data in continuous block to reduce disk I/O. 

Based on above discussions, we make the following contributions in the paper:

\begin{itemize}
	\item
	We study the schema integration problem for information integration on massive data sources. As we know, this paper studies the problem for the first time.
	
	\item
	We propose a framework of efficient and effective schema integration. Such framework could generate high-quality global schema within a limited cost. To support such framework, we use the knowledge base.
	
	\item
	To make our method suitable for a large amount of schemas, we design batch and incremental integration for different scenarios based on join algorithms. Such algorithms have benefits in both effectiveness and scalability for integration. For effectiveness, our algorithms consider both semantic and literal similarity between attribute names. It is suitable for various data schemas. For scalability, our algorithms are designed as external memory algorithms with the minimum disk I/O as the optimization goal.
	
	\item
	We conduct extensive experiments to verify the performance of the proposed methods. From the experimental results, our method can give a proper integrated schema. Also, our algorithm could integrate the large schema set efficiently by using small memory.
\end{itemize}

The remaining of this paper is organized as follows. Section~\ref{sec:preliminary} introduces preliminaries and backgrounds. Section~\ref{sec:overview} gives an overview of the whole framework. Section~\ref{sec:joinSchemaIntegration} introduces the join algorithms. Section~\ref{sec:batchIntegration} provides the detailed solution for batch integration.
Experimental results and analyses are given in Section~\ref{sec:experiment}. Section~\ref{sec:relatedWork} compares previous work and Section~\ref{sec:futureWork} concludes the paper.

\section{preliminary}
\label{sec:preliminary}
In this section, we introduce the backgrounds and definitions of the problem studied in this paper. At first, we give a brief introduction to knowledge base and edit distance. Then we define the problem and related symbols.

\subsection{Knowledge Base}
\label{sec:knowledgeBase}
The goal of involving knowledge base in our system is to measure the semantic similarity between attributes due to the synonyms and homonyms in attribute names.

Knowledge bases, such as \textit{Freebase}\footnote{\url{https://www.freebase.com/}}, \textit{WordNet}\footnote{\url{http://wordnet.princeton.edu/}}, \textit{Probase}\footnote{\url{http://research.microsoft.com/en-us/projects/probase/}} and \textit{YAGO}\footnote{\url{http://www.mpi-inf.mpg.de/departments/databases-and-information-systems/research/yago-naga/yago/}}, are often in graph structure with each concept as a node and each edge representing the relationship between concepts. Each concept refers to a real-world object, or a high-level concept consisting of objects. They can be attribute names of database schemas. Even though a knowledge base may have various structures, such classification does not lose generality. For examples, \textit{Freebase} has a two-level structure type-topic in a domain, both of which can be considered concepts. In a word, the structure of knowledge base is actually a graph $G$. Each node in the graph represents a concept.

Each node in the knowledge base is represented as a 3-tuple, (id, name, type). For example, the concept ``Pies'' and ``Sweet pies'' are represented as ($id1$, ``Pies'', ``wikicategory'') and ($id2$, ``Sweet pies'', ``wikicategory''), respectively. $id1$ and $id2$ are two ID numbers in the knowledge base, and ``wikicategory'' means a kind of knowledge got from wiki.

Some knowledge bases such as \textit{Freebase} have the ``is a'' form of structure, so the basic subclass relationship is represented as an edge in the knowledge base. Then a knowledge base contains a massive amount of such edges. Each edge in the knowledge base links two concepts with inclusion relationship.  Here, we show the definition of edge in the knowledge base and denote $S$ as the set of edges.

\newtheorem{defn}{Definition}
\label{def:edge}
\begin{defn}[edge]
	$\exists$ $a$,$b$ $\in$ $G$, if $a$ ``is a'' $b$, then edge ($a$,$b$) $\in$ $S$.
\end{defn}

For example, an edge (``Sweet pies'', ``pie'') means ``Sweet pies'' is a (kind of) ``pie''. Then the pair corresponds to an edge in $G$. We store all the edges in graph $G$ in a set $S$, so we can find concept relationship by scanning $S$. Depending on the circumstance of knowledge base, the amount of nodes in $G$ may be very large. Hence, the set $S$ is extremely complex with massive data.

To simplify the problem, when using $S$, we build a path as a more useful measurement of similarity based on the definition of edge. A path is from one concept to another, showing the relationship between them and consisting of a few edges. In such case, an edge is a special case of a path with length as 1. To measure the relationship within a path, we find the same tendency between similarity and path length. In general, the more similar two concepts are, the shorter the path between them is. As a qualitative connection between concepts, we will give more specific discussions in Section~\ref{sec:distanceFunction}.

However, such ``is a'' relationship is not the only way to define the similarity. Such as synonymy word base and \textit{NGD} (normalized Google distance) \cite{cilibrasi2006automatic}, other distance constraints can also achieve the same goal. With different definitions, we can obtain different results based on the distance. In this paper, we only discuss the algorithm based on knowledge base with ``is a'' relationship. Our proposed algorithm also work with other distance by preprocessing the knowledge base.

\subsection{Distance Function}
\label{sec:distanceFunction}
Based on the knowledge base, we define the semantic distance between two attributes in the schemas as follows.

\begin{defn}[semantic distance]
	\label{def:semanticDistance}
	$\exists$ $a$,$b$ $\in$ $G$, s.t. ($a$, $b$) $\in$ $S$, a semantic distance means the length of the path between $a$ and $b$, denoted as $dis_r (a, b)$.
\end{defn}

According to this definition, the smaller $dis_r$ is, the more similar $a$ and $b$ are, as described in Section~\ref{sec:knowledgeBase}. Then we use a threshold $\gamma$ to constrain whether two concepts are similar enough. Two concepts are regarded similar with the distance under the given threshold $\gamma$. For example, if we define $\gamma=2$ when $dis_r$ (``Sweet pies'', ``pie'')$=1$, then we regard ``Sweet pies'' and ``pie'' as related concepts.

With misspellings, an attribute name may not be found in the knowledge base. Thus, we should consider literal difference between attributes and concepts in the knowledge base. In this paper, we use edit distance~\cite{levenshtein1966binary}, a commonly-used distance function for strings to represent the literal distance between attribute names and concepts, denoted by $dis_t$. Utilization of edit distance will be discussed in detail in Section~\ref{sec:editDistance}.

With these considerations, we define following constraints of the determination whether attributes could be matched in schema integration.

\textbf{Distance Constraint}
\label{con:distance}
\begin{equation*}
\begin{aligned}
&dis(a,b)\le\varepsilon=\\
&\left\{
\begin{aligned}
&dis_r (c_a,c_b)\le\gamma \land dis_t (a,c_a)\le\varepsilon_t \land dis_t (b,c_b)\le\varepsilon_t,\\& \indent\exists c_a,c_b,dis_t (a,c_a)\le\varepsilon_t \land dis_t (b,c_b)\le\varepsilon_t\\
&dis_t (a,b)\le\varepsilon_t,\\ &\indent\forall c_a,c_b,dis_t (a,c_a)\ge\varepsilon_t \lor dis_t (b,c_b)\ge\varepsilon_t
\end{aligned}
\right.
\end{aligned}
\end{equation*}

For an attribute name $a$, if a concept $c_a$ in the knowledge base has the smallest literal distance with $a$ and such distance is smaller than the threshold $\varepsilon_t$, $c_a$ is considered to represent $a$. Thus, the semantic distance between two attributes $a$ and $b$ can be calculated according to corresponding concepts in the knowledge base. That is, $dis (a, b) \le \varepsilon$ means $dis_r (c_a, c_b) \le \gamma, dis_t (a, c_a) \le \varepsilon_t, dis_t (b, c_b) \le \varepsilon_t$. For example, we have two attributes $a$=``Sweet pies'' and $b$=``meet pie''. In the knowledge base, we have $c_a$=``Sweet pies'' and $c_b$=``meat pie''.
So $dis_t (a, c_a)$=0, and $dis_t (b, c_b)$=1. If $\varepsilon_t$=2, the literal distance between attributes and concepts is no more than the threshold, and then we should only check whether the semantic distance of corresponding concepts satisfies $\gamma$. In the knowledge base, ``Sweet pies'' and ``meat pie'' share the common neighbor concept ``pie'', so $dis_r (c_a, c_b)$ is 2. Since $dis (a, b)$ is within the threshold, ``Sweet pies'' and ``meat pie'' are treated as related attributes.

In the case that no literally similar concepts are found in the knowledge graph for either $a$ or $b$, we should only consider the direct relationship between the attributes. Then we use literal distance between them to measure their difference. For example, we misspell ``Abraham Lincoln'' as ``Abrehan Lincon'', and $dis_t$ between them is 3. If $\varepsilon_t$ is a number smaller than 3, then we could hardly find literally similar concepts in the knowledge base. Therefore, we take two attribute names ``Abrehan Lincon'' and ``Abraham Robinson'' as $a$ and $b$. Then the distance only depends on the literal difference between them. That is, $dis (a, b) = dis_t (a, b)$. For our example, $dis (a, b)$ is 6.

Note that even with the knowledge base and edit distance, attributes to be integrated could not be determined exactly due to the complexity in semantics. Take attributes ``import'' and ``export'' as an example. They are really literally similar and a path ``import''-``commodity''-``export'' in the knowledge base between them has length of 2, small enough. However, they are opposite to each other, not similar at all. Since semantic meaning is a difficult problem in schema matching\footnote{\url{http://en.wikipedia.org/wiki/Schema_matching}}, it may involve verification with human efforts. In this paper, we attempt to use automatic processing as much as possible. To achieve this goal, we should resolve false positives with further processing, which will be discussed in Section~\ref{sec:verifyJudging}.

\subsection{Edit Distance and Q-gram}
\label{sec:editDistance}
Edit distance is a measure to quantify how literally different two strings are\footnote{\url{http://en.wikipedia.org/wiki/Edit_distance}}. Many researchers have already worked on this problem, such as \cite{LiLu2009ed}, \cite{lin2011set} and \cite{xiao2008ed}. We use $dis_t (a, b)$ to represent the edit distance between $a$ and $b$. The threshold for edit distance is $\varepsilon_t$.

Current edit distance is mainly based on operating the q-gram structure of strings. A q-gram is a q-length substring of a string. Apparently, if edit distance between two strings is small, they should have many common q-grams. Here, we give out the common-used filtering condition as \cite{gravano2001using}.

\textbf{Count filtering} means that $a$ and $b$ must share at least $LB_{ab}$ common q-grams.
$$LB_{ab}=(\max(|a|,|b|)-q+1)-q*\varepsilon_t$$

Therefore, we can use $LB_{ab}$ as our constraint. We perform the count filtering on the alternative schemas and finally get those within the threshold $\varepsilon_t$.

Given an attribute, our approach first finds related concepts in the knowledge base. However, for schema integration on big data, it is impossible to scan all the concepts to find the exact ones. Here, we use an inverted list to organize the structure of the index.

For a word $w$, by using q-gram, it can be split up into $|w|-q+1$ parts. We name them $w_1$, $w_2$\ldots $w_k$ ($1 \le k \le |w|-q+1$). For $w_i$ ($1 \le i \le k$), it is in form of 2-tuple ($h_i$, $v_i$). $h_i$ is the hash value of string $w_i$, and $v_i$ is a set of words containing gram $w_i$. On the disk, we use $h_i$ as the index of $w_i$.

Then, the literal difference can be judged by count filtering here. Suppose that we match a string $s$ to a word set $W$. Then $s$ can be divided into q-grams $s_1$, $s_2$\ldots $s_k$ ($1 \le k \le |s|-q+1$).
By hashing $s_i$ ($1 \le i \le k$) and matching them to the attribute set $A$, we obtain the mapping ones $s_1$, $s_2$\ldots, $s_j$ ($j \le k$). Then we scan $v_1$, $v_2$\ldots $v_j$ and count the occurrence of the words.
If an attribute $a_i$ appears no less than $|s|-q+1-\varepsilon_t*q$ times, then we can consider that $s$ and $a_i$ are literally similar. We conflate them and get obtain integrated set.

\subsection{Problem Definition}
\label{sec:problemDefinition}
Let $\Sigma$ be a set of schemas. Each schema $s_i \in \Sigma$ is represented as a triple ($id_i$, $n_i$, $A_i$), where $id_i$ is an exclusive index that can be used to find the schema $s_i$ in $\Sigma$. $n_i$ is the name of the schema. $A_i$ is on behalf of the attributes of the schema. The attributes are the unit to integrate.

The problem of schema integration is to generate a global schema $S_g$ with attribute set $A_g$ from the schema set $\Sigma$. On one hand, all attributes in each schema $s_i \in \Sigma$ should be mapped to $S_g$. On the other hand, for each attribute $t \in A_i$ and its corresponding attribute in $A_g$ denoted by $a_t$, the distance between $t$ and $a_t$, denoted by $dis (t, a_t)$, should be smaller than a threshold. The definition of the distance function is just as discussed in Section~\ref{sec:distanceFunction}. With these considerations, the schema integration problem is defined as follows.

\textbf{Problem} [Schema Integration]
\label{prob:schemaIntegration}
Given a schema set $\Sigma$, generate a schema $S_g$ that satisfies that $a\in A_i$,  for $\forall s_i\in \Sigma$, $\exists a_t\in A_g$ with $dis (a, a_t)\le \varepsilon$, where $\varepsilon$ is the threshold.

For example, we have a schema containing attribute names such as ``Blackberry pie'', ``Strawberry pie'' and so on. During integration, we could find some other attributes  such as ``Savory pies'', ``Tiropita'', ``meat pie'' and ``tourtiere''. According to our definition, they are treated as similar and included into the integration answer.

\section{Overview}
\label{sec:overview}
In this section, we overview the solution of schema integration. As discussed in Section~\ref{sec:preliminary}, schema integration is based on the combination of semantic distance and literal distance to judge whether attributes are similar. For semantic distance generated from background knowledge, the first job is to initialize the knowledge base, which will be discussed in Section \ref{sec:initialization}.

The input of schema integration is a schema set $W$, a threshold $\varepsilon$ as well as the initialized knowledge base. Attributes in $W$ satisfying the distance constraint in Section~\ref{sec:distanceFunction} are considered similar and to be integrated. The result is a set $U$ with each element containing multiple attributes that are considered as similar.

For these scenarios, we design two algorithms for schema integration, batch integration and incremental integration. The former one is suitable for the cases that many schemas are to be integrated once. The latter one is for updating existing schemas by small-sized input. These will be introduced in Section~\ref{sec:overviewBatch} and Section~\ref{sec:overviewIncremental}, respectively.

For the convenience of processing, we develop a special data structure, cluster set, as the format of operands and output of following functions. Thus, without confusion, in the remaining part of this paper, we will not distinguish attribute set and cluster set.

\begin{defn}[cluster set]
	\label{def:clusterSet}
	With $S$ as the concepts set of the knowledge base, a cluster set is a set of pairs $\{U, S_U\}$ , where $U$ is a set of attributes and $S_U=\{(r, d)|d=min_{\forall t\in U}\\ \{dis(t, r)\} \land r \in S$\}. The function is the combination of both literal and semantic distance, as defined in Section~\ref{sec:preliminary}.
\end{defn}

\subsection{Initialization}
\label{sec:initialization}
As discussed in Section~\ref{sec:preliminary}, the knowledge base is used to measure the semantic similarity between attribute names. Knowledge bases often have complex structure and massive information. Our system uses just a small share of them. Thus, as the initialization, we extract the information required for the further steps in schema integration from the knowledge base. In this section, we discuss this step.

As shown in Section~\ref{sec:knowledgeBase}, in our system, the required knowledge is represented with ``is a'' relationship between concepts. Each concept is represented as a triple (id, name, type), where \textsf{id} is the index for concept, \textsf{name} is the identification string and \textsf{type} represents the part of knowledge base that the concept is from. Thus, the relationship between two concepts is described as a six-tuple (subId, subName, subType, superId, superName, superType).

\subsection{Batch Integration}
\label{sec:overviewBatch}
Batch integration integrates all schemas in batch. It is accomplished by clustering attributes in the schemas, and the attributes are merged into the integrated schema.

To achieve this goal, we develop two kinds of similarity join operations, ED Join and Semantic Join. They find pairs of attributes with edit distance smaller than a threshold and the semantic similarity larger than a threshold according to the knowledge base, respectively. Each pair in the join results are considered to be merged into the integrated schema.  After merging, the results need further processing due to transitivity issues.

Clearly, the cluster results should satisfy transitivity. That is, if $A$ and $B$ are similar attributes, B and C are similar attributes, then $A$ and $C$ should be similar. However, from the similarity join results, such transitivity may not be satisfied, since the similarity function does not satisfy the transitivity. To solve this problem, we develop a further step, i.e. resolve.

The pseudo code of batch integration is shown in Algorithm~\ref{alg:batchIntegration}. In this algorithm, firstly, all the attributes in input schemas are added to a set $U$ (Lines 1-3).  Then, to compress input set, we perform ED Join on $U$ to merge all the literally similar attributes (Line 4). Next, we perform Semantic Join on $U$ to merge all semantically similar attributes (Line 5). After this step, all attribute pairs in the results are considered as a cluster. As discussed above, the transitivity problem may occur in the results. To solve the transitivity problem, we verify the cluster generated by the two join operations. This task is accomplished in Resolve($U$)(Line 6), whose details will be discussed in Section~\ref{sec:resolve}.

\begin{algorithm}
	\caption{Batch Integration}
	\label{alg:batchIntegration}
	\scriptsize
	\KwIn{schema batch $W$}
	\KwOut{integration set $U$}
	\ForEach{$w \in W$}
	{
		$U \leftarrow U \cup A_w$\;
	}
	$U \leftarrow U - $ EDJoin($U$,$U$)\;
	$U \leftarrow U - $ SemanticJoin($U$,$S$)\;
	$U \leftarrow$ Resolve($U$)\;
	\Return $U$\;
\end{algorithm}

\subsection{Incremental Integration}
\label{sec:overviewIncremental}
Different from batch integration, incremental integration integrates schemas to the existing global schema one by one. Such approach is suitable for adding data sources.

To reduce the cost of integrating a schema to the global schema by both literal and semantic matching, we maintain a cluster set $U$ containing all attributes in the global schema. When we add a new schema $K$, for each attribute $a$ in $K$, if $a$'s literally and semantically similar attributes are not found in $U$ then $a$ is inserted into $U$, and the cluster set is updated according to the new updated attributes. To avoid false positive, verification phase is also adopted.

The pseudo code for incremental integration is shown in Algorithm~\ref{alg:incrementalIntegration}. Firstly, as discussed, $K$ is joined with the maintained attribute set $U$ (Line 1). Then, the results are verified in Line 2. The attributes in $K$ that is not matched with any attributes in $U$ are collected in set $V$ (Line 3). Attributes in $V$ and related concepts in the knowledge base $S$ are added to $U$ (Line 4-6).

\begin{algorithm}
	\caption{Incremental Integration}
	\label{alg:incrementalIntegration}
	\scriptsize
	\KwIn{inserting schema $K$, integration set $U$}
	\KwOut{integration set $U'$ after insert}
	$T \leftarrow$  EDJoin($K$,$U$)\;
	$R \leftarrow$  Verify($T$,$U$)\;
	$V \leftarrow K-R$\;
	$V \leftarrow$ EDJoin($V$,$S$)\;
	$V \leftarrow$ SemanticJoin($V$,$S$)\;
	$U \leftarrow U \cup V$\;
	$U \leftarrow$ Resolve($U$)\;
	\Return $U$\;
\end{algorithm}

We have existing integration results and one inserting schema as the input of incremental integration. First, we perform ED Join on adding schema with existing results. To avoid false positive, we also proceed verification process in Section~\ref{sec:verifyJudging}. As it is confirmed, we add it into the integration set. To make the integration set suitable for following insertion, we add literally and semantically similar attributes into $S_U$ of the cluster set. In this way, we can judge whether future added attributes is similar to some attributes in this set easily.

From above discussions, these two schema integration algorithms share three common operations, (1) ED Join, the similarity join based on edit distance; (2) Semantic Join, semantic similarity join based on knowledge base, and (3) Resolve, the verification and partition of clusters. Note that function Verify() in Algorithm~\ref{alg:incrementalIntegration} is a part of Resolve() function in Algorithm~\ref{alg:batchIntegration}.

In the following sections, we first introduce ED Join, Semantic Join and Resolve in Section~\ref{sec:edJoin}, Section~\ref{sec:semanticJoin} and Section~\ref{sec:resolve}, respectively. Even based on these operations, batch integration is not straightforward and will be discussed Section~\ref{sec:batchIntegration} in detail.

\subsection{Verification}
\label{sec:verifyJudging}
Due to the work flow of ED Join, non-related words with small spelling difference can be joined together such as ``works'' and ``words''. For Semantic Join, a large threshold may lead to integrating non-related concepts in the knowledge base. Therefore, false positive may be involved in the answer set. In order to eliminate the false positive, we propose the verification approach, which has two parts, value verification and manual verification.



\noindent \underline{Value Verification} In the integration problem in Section~\ref{sec:problemDefinition}, we perform integration only based on the attribute names in schemas. Value verification aims to verify the results according to the values of the attributes. Such approach is based on the observation that if two attributes are similar, values of them should be same or similar as well. Therefore, values of the attributes can be used to judge the relationship between attribute names and to correct false positives. To find the relationship of attribute values, structural analysis is a simple but effective way. We develop some rules as the judgment standard. The discovery approaches of more rules are left for further research.

\begin{itemize}
	\item \textbf{Type} Each attribute has its data type, such as integer, string, list and so on. Data values sharing the same data type are possibly similar, especially for some complex structure. For example, if attributes contain string sets of 11 people's names, they can be treated as similar attributes as football team name list. Also, in contrast, if values in some attributes are strings, while those in others are integers, they can be unlikely similar. Thus, we use the type as the first judgment rule.
	\item \textbf{Affix} Prefix and suffix can be a specific word structure to help as well. For example, if values of attributes share the same prefix or suffix such as ``\$$\dots$'', they can be the cost or money record. Thus, the prefix or suffix are used as the second judgment rule.
\end{itemize}

Above rules are used to judge false positives. If attributes judged similar by former steps obey these rules of values, they should be judged as false positives and the relationship are denied.

\noindent \underline{Manual Verification} Even though value verification is effective in some cases, it is a weak constraint and sometimes unavailable. Generally, it is difficult to check attributes without distinct structure or some attributes without values. For more accurate integration, we involve manual efforts for further verification. Thus, crowdsourcing is adopted on some small, accuracy needed field to improve the accuracy.

\section{JOIN SCHEMA INTEGRATION}
\label{sec:joinSchemaIntegration}
In this section, we propose ED Join and Semantic Join algorithms. Both of these two operators are necessary in the schema integration. The implementation of them are different. ED Join attempts to find the pairs of strings with edit distance smaller than a threshold, while Semantic Join finds the pairs of concepts on the knowledge graph with distance smaller than a threshold. The details of their implementations will be discussed later in Section~\ref{sec:edJoin} and Section~\ref{sec:semanticJoin}, respectively.

According to the definition of cluster set, the operators of ED Join and Semantic Join are defined as follows.

\begin{defn}[ed join]
	\label{def:edJoin}
	Given two families of cluster sets, $R$ and $T$, and a threshold $d$, two elements ($U_1$, $S_1$) and ($U_2$, $S_2$) from $R$ and $T$, respectively, are ED joined if they satisfy one of the following constraints.
	\begin{enumerate}
		\item $\min\limits_{r_1\in U_1,r_2\in U_2} dis_t (r_1,r_2)\le \varepsilon_t$
		\item $\exists(r,d)\in S_2, \min\limits_{r_1\in U_1} dis_t (r_1,r)\le \varepsilon_t-d$
		\item $\exists(r,d)\in S_1, \min\limits_{r_2\in U_2} dis_t (r_2,r)\le \varepsilon_t-d$
	\end{enumerate}
	The ED Join result of ($U_1$, $S_1$) and ($U_2$, $S_2$) is a pair ($U$, $S_U$), where $U=U_1\cup U_2$ and $S_U=\{(r, d)|r\in S \land d=\min_{t\in U} \{dis(r, t)\}\}$.
\end{defn}

\begin{defn}[semantic join]
	\label{def:semanticJoin}
	Given two families of \\cluster sets $R$, $T$, and a threshold $d$, two elements ($U_1$, $S_1$) and ($U_2$, $S_2$) are from $R$ and $T$, respectively are semantically joined if they satisfy one of the following constraints.
	\begin{enumerate}
		\item $\min\limits_{r_1\in U_1,r_2\in U_2} dis_r (r_1,r_2)\le \gamma$
		\item $\exists(r,d)\in S_2, \min\limits_{r_1\in U_1} dis_r (r_1,r)\le \gamma-d$
		\item $\exists(r,d)\in S_1, \min\limits_{r_2\in U_2} dis_r (r_2,r)\le \gamma-d$
	\end{enumerate}
	The result of Semantic join on ($U_1$, $S_1$) and ($U_2$, $S_2$) is a pair ($U$, $S_U$), where $U=U_1\cup U_2$ and $S_U=\{(r, d)|r\in S \land d=\min_{t\in U} \{dis(r, t)\}\}$.
\end{defn}

ED Join joins attributes with edit distance within a given threshold while Semantic join is for semantically similarity. Initially, sets are joined according to the similarity between $r_1 \in U_1$ and $r_2 \in U_2$. Such direct relationship between attributes is described as the first constraints in Definition~\ref{def:edJoin} and Definition~\ref{def:semanticJoin}. For two attributes $r_1$ in $U_1$ and $r_2$ in $U_2$, if the distance between them is no more than one of the thresholds, following condition 1 they are regarded similar and the cluster sets are able to be integrated.
For example, we have two attributes of $r_1$ and $r_2$, ``Sander'' and ``Sunder''. Of course, they are literally similar since $dis_t$ is 1. Thus, $U_1$ and $U_2$ can be joined.

Also, corresponding to the definition of cluster set, Definition~\ref{def:edJoin} and Definition~\ref{def:semanticJoin} also involve condition 2 and 3. That is, if the distance between $r_1$ in $U_1$ and $r$ in $S_2$ is within $\gamma-d$, we can regard $r_1$ and $r$ are similar and two cluster set where $r_1$ and $r$ are from can be joined. Such judgement process works for both ED Join and Semantic Join.

Both of ED Join and Semantic Join require to combine two cluster sets ($U_1$, $S_1$) and ($U_2$, $S_2$) into one ($U$, $S_U$). We define this operator as \textit{Pair Join}.
We show the pseudo code for the implementation of Pair Join of two single cluster sets in Algorithm~\ref{alg:pairJoin}. Firstly, we union two $U$ sets of these two cluster sets (Line 1). Each attribute in $S_1$ or $S_2$ should be included in $S$ as well.
However, during the join, the parameter $d$ of ($r$, $d$) in $S_U$ should be updated. For one pair ($r$, $d$), if one $v$ in $U$ satisfies that $dis(r,v)$ is smaller, $d$ should be adjusted. Hence, we examine and update the pairs (Lines 2-9).

\begin{algorithm}
	\caption{Pair Join}
	\label{alg:pairJoin}
	\scriptsize
	\KwIn{two cluster pairs ($U_1$, $S_1$) and ($U_2$, $S_2$)}
	\KwOut{joined pair ($U$, $S$)}
	$U \leftarrow U_1 \cup U_2$\;
	\ForEach{$(r,d)\in S_1 \cup S_2$}
	{
		\If{$dis_t(r,v) \le d$}
		{
			$S \leftarrow S \cup (r,dis_t(r,v))$\;
		}
		\Else
		{
			$S \leftarrow S \cup (r,d)$\;
		}
	}
	\Return ($U$,$S$)\;
\end{algorithm}

Based on the Pair Join solution, ED Join and Semantic Join are two crucial steps to finish batch integration and incremental integration. We will discuss ED Join and Semantic Join respectively in Section~\ref{sec:edJoin} and Section~\ref{sec:semanticJoin}.

\subsection{ED Join}
\label{sec:edJoin}
ED Join joins cluster sets with literally similar attributes. It is similar as the similarity join on string sets based on the edit distance \cite{levenshtein1966binary,LiLu2009ed}. As an efficient approach of similarity joins on string sets, we adapt q-gram-based methods for ED Join.

As the basic data structure, we use inverted list with each q-gram as an entry. The attributes in the cluster sets are indexed with q-grams, respectively. With input denoted as $R$ and $T$, the q-gram-based inverted lists for attribute sets of them are $X_R$, $X_T$ for $U$ and $Z_R$, $Z_T$ for $S_U$, respectively. Since in the ED Join, two cluster sets could be joined according to three constraints in Definition 4, then the q-gram-based similarity join is performed according to index pairs $X_R$ and $X_T$, $X_R$ and $Z_T$, as well as $X_T$ and $Z_R$, respectively.

As we all know, there are always some mistakes in the knowledge base~\cite{fayyad1996data}, even more common in the words of attributes. To decrease the negative impact, fault tolerance mechanism of misspelling is in demand. Adapting inverted list in ED Join, join between all literally similar attributes such as misspelling words can be accomplished over sets at the same time. The inverted lists of clusters are generated offline and stored on the disk for reuse.

The pseudo code for ED Join algorithm is shown in Algorithm~\ref{alg:edJoin}. First, the inverted list for $q$-grams is constructed for $R$ and $T$ (Lines 1-4). Then, we use function ED Merge to perform similarity join based on the q-gram list indices according to the three constraints in Line 5-7.

In the ED Merge function, we have an input $H$ as the index list, and output $K$ as a set of joined pairs. In the list $H$, we denote each gram as $g$, while each $g$ is followed by a list \{$v_1, v_2, \dots$\} as the attributes that contain $g$.
First, we count the appearance times of each attribute $v$ in each part of the list $H$ (such as $X_R$ $\cap$ $X_T$ in Line 5) (Line 10), and initialize the answer set $K$ as an empty set (Line 11).
Then, for each $v$ appearing more than $|v|-q+1-\varepsilon_t*q$ times, as mentioned in Section~\ref{sec:editDistance}, there exists attributes similar to $v$ in $H$. And then the cluster set where $v$ is from should be joined into $K$ using Pair Join (Lines 12-16). Locate($v$) just returns the cluster set that $v$ belongs to. Finally, we obtain the answer cluster set $M$ which contains integrated attributes for ED Join.

\begin{algorithm}
	\caption{ED Join}
	\label{alg:edJoin}
	\scriptsize
	\KwIn{two cluster sets $R$ and $T$, threshold $\varepsilon_t$ and $d$}
	\KwOut{joined cluster sets $M$ including pairs ($U$, $S$)}
	$X_R \leftarrow$  q-gram($R.U_i$)\;
	$X_T \leftarrow$  q-gram($T.U_i$)\;
	$Z_R \leftarrow$  q-gram($R.S_i$)\;
	$Z_T \leftarrow$  q-gram($T.S_i$)\;
	$M \leftarrow M \cup$ EDMerge ($X_R \cap X_T$)\;
	$M \leftarrow M \cup$ EDMerge ($X_R \cap Z_T$)\;
	$M \leftarrow M \cup$ EDMerge ($X_T \cap Z_R$)\;
	\Return $M$\;
	\SetKwBlock{Begin}{function EDMerge}{end}
	\Begin{
		\KwIn{q-gram $H$}
		\KwOut{set of joined pairs $K$}
		Count($v\in g\in H$)\;
		$K \leftarrow \emptyset$\;
		\ForEach{$v\in g\in H$}
		{
			\If{$count[v] \ge |v|-q+1-\varepsilon_t*q$}
			{
				$K \leftarrow$ PairJoin($K$, Locate($v$))\;
			}
		}
		\Return $K$\;
	}
\end{algorithm}

In this part, the time complexity is mainly subject to $|R|$ and $|T|$, due to their large size. Therefore, the major cost of this join is in the step of q-gram denoted as O($|R|$+$|T|$). Since $X_R \cap X_T$ is no more than $R$ or $T$, the costs of Line 10 and Lines 12-16 are both O($|H|$). Thus, the time complexity of EDMerge is O($|X_R \cap X_T|$), no more than O($|R|$+$|T|$). In conclusion, the time complexity is O($|R|$+$|T|$).

\subsection{Semantic Join}
\label{sec:semanticJoin}
As mentioned in Section~\ref{sec:distanceFunction}, more concepts can be regarded similar within the threshold $\gamma$. However, such goal is much difficult to achieve. Under a given threshold, the integration set may form a circle-like subgraph in the knowledge base. The input attributes are the centers of the subgraph, and all the similar attributes to be integrated are in the subgraph. When the knowledge base is too large, due to the large cost of time and space, such easy idea is not feasible and following problems arise. How to locate attributes of the input set on the knowledge graph sufficiently? How to find related concepts efficiently on the knowledge graph? How to perform the merge when overlapping among different-source circle happens? To ensure the efficiency, we first show how we store the knowledge base on the disk by hash in Section~\ref{sec:hash}.
The main algorithm of Semantic Join is detailed discussed in Section~\ref{sec:joinAlgorithm}. To improve the implementation of the algorithm, we propose some techniques in Section~\ref{sec:implementations}.

\subsubsection{Hash-based Storage}
\label{sec:hash}
Due to the large size of knowledge base, it has to be stored on the disk. To ensure the performance of disk-based algorithm, we develop index in this section. At first, we discuss the operation that requires to access the disk, and then propose the index structure and its applications. As discussed in Section~\ref{sec:overview}, Semantic Join finds the concept pairs with path smaller than a threshold in the knowledge graph. A path is represented by triple ($start$, $end$, $len$), where $start$, $end$ and $len$ are the start concept, end concept and the length of the path, respectively. To generate the path, the basic operation is to find $k$-hop neighbors for a start concept $start$, which are all concepts having a path from $start$ with the length within $k$. Here, we generate a hash table of concepts in the knowledge base, and then Semantic Join is converted to a series hash join with the support of hash.

To accelerate join processing, we maintain a hash table for $k$-hop neighbors of all concepts, denoted as $H_k$. Such table is used to find required pre-processed neighbors relationship within O(1) time complexity. Thus, running time is saved by turning the paths in the knowledge base into accessing a series of segments in hash tables. We define neighbor table as first.

\begin{defn}[neighbor table]
	\label{def:hashTable}
	$t$ is an attribute and $P$ is the set of all paths in the knowledge base. $H_k(t)$ is a table on the disk indexed by hash value of string $t$, s.t.
	$$
	H_k(t)=\{a_i|(t, a_i, d)\in P\land d=k\}
	$$
\end{defn}

The neighbor table $H_k(t)$ accepts one concept $t$ and returns all the concepts having a $k$-length path with it. It is a hash list on the disk. Adapting such hash structure, we can search $k$-length path within constant time. However, it is not wisdom to construct a table with a large $k$, since when $k$ gets large, the number of neighbors increases significantly which makes the pre-process cost too much. Also, a long distance path is hardly to use because it is meaningful only when the distance is under a certain value. In this way, we choose to store 1-hop concepts, which are defined as $H_1$. Note that neighbor table $H_1$ just express the information of edges.

We also perform further optimization for such hash-based storage. The common goal of a hash function is to choose a suitable hash function and can reduce the conflict between concepts. However, based on the problem stated in Section~\ref{sec:problemDefinition}, we try to do the opposite thing. That is, by setting a proper hash function, we locate the elements in one cluster set into the same bucket which is in consecutive blocks on the disk.



Therefore, the aim of the hash function is to congregate attributes close on the disk. Here, we separate the table into several buckets. Attributes that will be accessed together are located in the same bucket. For example, attributes \{$a_1,a_2\dots a_i$\} in one cluster set should be included in one hash bucket, then the disk accessing time for the join operation is reduced. The hash number of each item is made up by bucket number and offset in the bucket. Items in the same bucket share the same bucket number. For example, we choose a hash seed as 13, the bucket length as 10,000 and the base offset of this bucket as 1,000,000. The offset of each attributes is shown in Table~\ref{tab:bucketHash}.

\begin{table}
	\scriptsize
	\centering
	\caption{Example of Bucket Hash Offset}
	\begin{tabular}{ccc}
		\hline
		Attribute &	Offset in Bucket &	Total Offset\\
		\hline
		Name & 9277 &	1009277\\
		\hline
		Speed &	5109 &	1005109\\
		\hline
		Amount &	2380 &	1002380\\
		\hline
		Streetname &	2708 &	1002708\\
		\hline
	\end{tabular}
	\label{tab:bucketHash}
\end{table}

The algorithm is shown in Algorithm~\ref{alg:bucketHash}.

\begin{algorithm}
	\caption{Bucket Hash}
	\label{alg:bucketHash}
	\scriptsize
	\KwIn{one set of attributes of $A$, offset base value $R_0$}
	\KwOut{set $K$ of (hashkey, attribute)}
	$K \leftarrow \emptyset$\;
	\ForEach{$a \in A$}
	{
		$k \leftarrow 0$\;
		\ForEach{$s \in a$}
		{
			$k \leftarrow (k * hash\_seed + s)$\;
			$s \leftarrow next(s)$\;
		}
		$k \leftarrow k \% bucket\_length$\;
		$k \leftarrow k + R_0$\;
		$K \leftarrow (k, a)$\;
	}
	\Return $K$\;
\end{algorithm}

In this algorithm, we take a set of attributes $A$ to be aggregated on the disk as input, together with the base offset $R_0$ where the bucket starting on the disk. For each attribute $a$, we calculate its hash value $k$ (Lines 2-7). Since these attributes are supposed to be placed in one bucket to reduce disk accessing time, the offset of each attribute is calculated by adding bucket offset $k$ to base offset $R_0$ (Lines 8-9). Then all pairs of hash key and attribute are added into set $K$ as the output.

Hash buckets are used for the join algorithm in Section~\ref{sec:joinAlgorithm}. Such table is generated at the beginning of the integration by offline processing. When the threshold gets larger, the pre-computation for longer paths may be required to accelerate the join. 


\subsubsection{Join Algorithm}
\label{sec:joinAlgorithm}
The idea of Semantic Join algorithm is to aggregate all the semantic-related attributes. The target attribute set $R$ is joined with edges in the knowledge base denoted as $E$. Under the semantic threshold $\gamma$, the process of Semantic Join is shown as follows logically.
$$
(R) \cup (R\Join E) \cup (R\Join^2 E) \cup \dots \cup (R\Join^{\gamma-1} E) \cup (R\Join^\gamma E)
$$

During the join above on the knowledge base, we link paths in various lengths. Each linking needs matching between path end nodes and path start nodes. To accelerate computation, we design path set, a sophisticated data structure, which aggregates paths with the same end node in the same hash bucket. Such data structure is defined as follows.


\begin{defn}[path set]
	\label{def:pathSet}
	$P_a$ is a path set, all paths in which share the same end node $a$, s.t.
	$$P_a = \{(start, k) | \exists start \in H_k(a)\}$$
\end{defn}

Based on such data structure, the pseudo code of Semantic Join is proposed in Algorithm~\ref{alg:semanticJoin}. The algorithm takes the target cluster set $R$, the threshold $\gamma$ and the knowledge base with hash-based storage as input, and the output is the joined cluster sets $M$. The algorithm can be divided into three steps as follows.

\begin{enumerate}
	\item \textbf{Initialization(Lines 1-6)} This step scans the input attributes, and collects all their 1-hop neighbor concept information.
	
	\item \textbf{Path expanding(Lines 7-10)} From the nearest neighbor, in this step, we perform the join on the knowledge base and obtain input's similar concepts.
	\item \textbf{Cluster set merging(Lines 11-20)} This step merges cluster sets with similar attributes.
\end{enumerate}

\begin{algorithm}
	\caption{Semantic Join}
	\label{alg:semanticJoin}
	\scriptsize
	\KwIn{one cluster set $R$, semantic threshold $\gamma$ and 1-hop neighbor table $H_1$}
	\KwOut{joined cluster sets $M$}
	$P \leftarrow \emptyset$\;
	$M \leftarrow \emptyset$\;
	\ForEach{$w \in U_i |(U_i,S_i)\in R$}
	{
		$P_h \leftarrow P_h \cup \{(w,1)|h \in H_1(w)\}$\;
		$P \leftarrow P \cup P_h$\;
	}
	\For{$i$=1 \KwTo $\gamma -1$}
	{
		\ForEach{$P_i \in P$}
		{
			$P_j \leftarrow P_j \cup \{(start,len+1)|j \in H_1(i)\}$\;
			$P \leftarrow P \cup P_j$\;
			\If{$j \in U_i|(U_i,S_i)\in R$}
			{
				\ForEach{$start \in P_j$}
				{
					$M \leftarrow M \cup PairJoin(atCluster(start),(U_i,S_i))$\;
				}
			}
			\If{$(j,d) \in S_i|(U_i,S_i)\in R \land len+1+d< \gamma$}
			{
				\ForEach{$start \in P_j$}
				{
					$M \leftarrow M \cup PairJoin(atCluster(start),(U_i,S_i))$\;
				}
			}
		}
	}
	\Return $M$\;
\end{algorithm}

In the algorithm, we use two variables, $P$ containing path sets, and $M$ as the collection of joined cluster sets (Lines 1-2). For initialization, we locate all the concepts to be integrated (center concept) in the knowledge base. Then, 1-hop neighbors of all center concepts in $R$ are obtained and putted in $P$ grouped by their end nodes (Line 3-6). After that, multi-hop neighbors are obtained. The details will be discussed later.

We use an example to illustrate the algorithm. As for one concept ``Blackberry pie'' in the cluster set, the corresponding content in $H_1$(``Blackberry pie'') is \{``American pies'', ``Sweet pies''\}. After adding them into $P$, the set $P$ is shown in Table~\ref{tab:initializing}.

\newcommand{\tabincell}[2]{\begin{tabular}{@{}#1@{}}#2\end{tabular}}

\begin{table}[h]
	\scriptsize
	\centering
	\caption{Initial Set $L$}
	\begin{tabular}{ccc}
		\hline
		(start, len) & end\\
		\hline
		\{(Blackberry pie,1)\} &	American pies\\
		\hline
		\{(Blackberry pie,1)\} &	Sweet pies\\
		\hline
	\end{tabular}
	\label{tab:initializing}
\end{table}

After locating the target, we just link $w$ to the knowledge base. With the paths directly linked to $w$, we expand paths to obtain more concepts related to it. However, the knowledge base may contain too many concepts, most of which never have a relationship with $w$. For example, we regard ``Blackberry pie'' as our target. Considering concepts ``song'' or ``computer'', there is no reason to put them together since they are non-related at all. In contrast, concepts ``Strawberry pie'', ``Tiropita'' or some other pastry are reasonably integrated according to human ideas, since they are in the same class of ``Blackberry pie''.

As we know, too many concepts may be contained in the knowledge base. If we scan each concept pair ($w_1, w_2$) to judge the relationship, it is costly due to the nested loop. Also, most of the searching is meaningless, since a large amount of concepts in the knowledge base have no relationship with the target. Motivated by this, we attempt to focus on given concepts and expand around $w$. Our duty then turns to find concepts in a subgraph with paths whose length is no more than the threshold.

After adding 1-hop neighbor information into $P$ in step 2, we perform join with $H_1$ to expand from the end concepts of each path set.

Then, as for algorithm Semantic Join, the expanding step is based on $P$ obtained from the initialization step. Based on 1-path set to get $\gamma$-path subgraph, times of expanding loop iteration is $\gamma -1$ (Line 7). For each path set $P_i$ in set $P$, the first thing is to find paths that end concept $i$ can link to. By using 1-hop neighbor table $H_1$, all the paths ($i$, $j$, 1) can be filtrated out and linked with $i$, then the paths expand to $j$. And then, we insert ($start, len+1$) into $P_j$ (Lines 8-10).

Considering the motivating example, a piece of $H_1$ is shown in Table~\ref{tab:aPieceH}. After operation in Lines 8-10, each path set $P_i$ in $P$ now contains paths no longer than $\gamma$ as shown in Table~\ref{tab:expanedPathList}.
It means that we have found the relationship from ``Blackberry pie'' to other similar concepts under threshold $\gamma$, such as (``Blackberry pie'', ``pie''), (``Blackberry pie'', ``Strawberry pie''). Therefore, the end concepts in $P$ are the similar ones to be integrated with ``Blackberry pie''.

\begin{table}
	\scriptsize
	\centering
	\caption{A Piece of $H_1$}
	\begin{tabular}{ccc}
		\hline
		start &	end &	len\\
		\hline
		American pies & Key lime pie &	1\\
		\hline
		American pies &	pie &	1\\
		\hline
		American pies &	Natchitoches meat pie &	1\\
		\hline
		Sweet pies &	Strawberry pie &	1\\
		\hline
		Sweet pies &	pie &	1\\
		\hline
		pies & Savoury pie & 1\\
		\hline
		Natchitoches meat pie & Savoury pies & 1\\
		\hline
		Savoury pies & Tiropita &	1\\
		\hline
	\end{tabular}
	\label{tab:aPieceH}
\end{table}

\begin{table}
	\scriptsize
	\centering
	\caption{Path List Set $L$ after Expanding}
	\begin{tabular}{ccc}
		\hline
		(start, len) & end\\
		\hline
		\{(Blackberry pie,1)\} &	American pies\\
		\hline
		\{(Blackberry pie,1)\} &	Sweet pies\\
		\hline
		\{(Blackberry pie,2))\} &	Key lime pie\\
		\hline
		\{(Blackberry pie,2)\} &	pie\\
		\hline
		\{(Blackberry pie,2), (Blackberry pie,4\} &	Natchitoches meat pie\\
		\hline
		\{(Blackberry pie,2)\} &	Strawberry pie\\
		\hline
		\{(Blackberry pie,3)\} &	Savoury pies\\
		\hline
		\{(Blackberry pie,4)\} &	Tiropita\\
		\hline
	\end{tabular}
	\label{tab:expanedPathList}
\end{table}

When executing $P_j \leftarrow P_j \cup \{(start,len+1)|j \in H_1(i)\}$ (Line 9) to expand the paths on the knowledge base, $P_j$ is updated by inserting new paths. However, $P_j$ often contains duplicated concepts with the same start and end concepts, such as the set \{(Blackberry pie,2), (Blackberry pie,4\} in Table~\ref{tab:expanedPathList}. To increase the quality of the results of Semantic Join, such duplication should be eliminated. We use an example to show the generation of the duplication.

From ``Blackberry pie'' to ``Natchitoches meat pie'', we may construct it in the process of ``Blackberry pie''-``American pies''-``Natchitoches meat pie''. The path is (``Blackberry pie'', ``Natchitoches meat pie'', 2). However, ``Blackberry pie'' to ``Natchitoches meat pie'' can also be constructed as ``Blackberry pie''-``American pies''-``pie''-``Savoury pie''-``Natchitoches meat pie''. The path is (``Blackberry pie'', ``Natchitoches meat pie'', 4). Both of these two paths are from ``Blackberry pie'' to ``Natchitoches meat pie'' and they are duplicated.

According to our definition of semantic distance in Definition~\ref{def:semanticDistance}, the difference between two concepts is represented as the length in the algorithm. Therefore, the smaller the length is, the more similar the two concepts are. Under a given threshold $\gamma$, a shorter path (length = $d$) have more chance to join ($\gamma -d$ times which is larger) with concepts. For paths sharing the same start and end concept, the shorter one can collect more similar concepts by join. Therefore, as for the example above, the path with length of 2 is a better choice for the integration. In conclusion, while executing $P_j \leftarrow P_j \cup \{(start,len+1)|j \in H_1(i)\}$ (Line 9), if a pair ($start,len$) in $P_j$ has the same start concepts, only the path with the shortest length is kept in $P_j$.

Finally, the end concept $j$ in Line 9 is used to judge whether cluster sets $(U_i, S_i)$ should be merged. If the end concept $j$ is in any $U_i$, then the cluster set where $start$ from, $(atCluster(start))$, is joined with $(U_i, S_i)$ and then added into $M$ (Lines 11-15). Also, if the end concept $j$ is in any $S_i$ with distance $d$ to a corresponding concept in $U_i$, satisfying $len+1+d<\gamma$, which means the distance between $start$ and any concept in $U_i$ is no more than $\gamma$, then the cluster set $(atCluster(start))$ is also joined with $(U_i, S_i)$ and then added into $M$ (Lines 16-20).

In this part, the time complexity is mainly affected by $|R|$ and $|P|$. At Lines 3-5, $P$ is initialized by elements in $R$ by the complexity of O($|R|$). Lines 7-22 is a dual loop under range of $\gamma$ and $|P|$. In each iteration, hash finding or set union can be implemented under O(1). While Pair Join has the complexity of O($|C|$), by denoting the average length of cluster set as $C$. Then the cost of the loop is O($\gamma |P|\cdot|C|$). Hence, the total complexity is O($|R|+\gamma |C|\cdot|P|$).

\subsubsection{Implementations}
\label{sec:implementations}
The application of the neighbor table mentioned in Section~\ref{sec:hash} can improve the efficiency of Semantic Join. However, when the threshold gets larger, $H_1$ is scanned $|\gamma| - 1$ times and the cost is larger. Based on the definition of neighbor table, the concepts found by $H_k$ and $H_{k_1}\Join H_{k_2}\Join \dots \Join H_{k_m} (k_1+k_2+\dots+k_m=k)$ are the same. Therefore, the main idea is to use neighbor table with higher $k$ to diminish the accessing times of $H_{k_i}$. Also, the generation of the neighbor table is costly when $k$ is high, so it is infeasible to construct a neighbor table with every $k$.
Here, we use integer power of 2, i.e. 1, 2, 4, 8, 16, 32, 64, $\cdots$, to balance initialization cost and combine the threshold of join algorithm by adding up some of these numbers. That is, we choose to generate $H_k$ which has $k$ as exponents of two, such as $H_1$, $H_2$, $H_4$ and so on. For example, we have the threshold as $\gamma=6$, whose binary representation is 110. From the binary representation, we know that 6 is the sum of 4 and 2, which means that $H_4$ and $H_2$ are enough to accomplish the task. Such approach has the following advantages.

\begin{itemize}
	\item
	The sum of integer power of 2 can cover each integer with a few base numbers. In our problem, this advantage means that we can generate and store fewer neighbor tables but satisfy threshold.
	\item
	An integer power of 2 is easy to calculated on computers by bit shifting. These bit-wise operations is light during joining. Also, we can use mask bit to improve the efficiency of neighbor table selection.
	
\end{itemize}

\noindent \underline{Time Complexity Analysis}
Based on proposed ED Join and Semantic Join algorithms, we analyze the complexity of batch integration and incremental integration.

\noindent \textbf{Batch Integration}
Following the flow in Algorithm~\ref{alg:batchIntegration}, its time complexity is as follows, where the average length of a cluster set is denoted by $C$. The step of searching similar neighbor concepts is optimized by pre-hash process.

\begin{eqnarray*}
	&&O(Batch Integration)\nonumber\\
	&=&O(|U|)+O(EdJoin(U,U))+O(SemanticJoin(U)\nonumber\\
	&=&O(|U|)+O(|U|+|U|)+O(\gamma |U|\cdot|C|))\nonumber\\
	&=&O(\gamma |U|\cdot|C|)\nonumber\\
\end{eqnarray*}

\noindent \textbf{Incremental Integration}
Following the flow in Algorithm~\ref{alg:incrementalIntegration}, with input as inserting existing integration set $U$ and schema $K$ (much smaller than $U$ according to the problem), the time complexity is as follows. The step of searching similar neighbor concepts is optimized by pre-hash process.

\begin{eqnarray*}
	&&O(Incremental Integration)\nonumber\\
	&=&O(EdJoin(K,U)+O(EdJoin(V,S))\nonumber\\
	&&+O(SemanticJoin(V,S))\nonumber\\
	&=&O(|K|+|U|)+O(|K|+|S|)+O(\gamma |K|\cdot|C|)\nonumber\\
	&=&O(|U|)+O(\gamma |K|\cdot|C|)\nonumber\\
\end{eqnarray*}

According to the analysis above, the time cost of the proposed algorithm is unrelated to the size of knowledge base. Without accessing the knowledge base too many times, the algorithm saves much time and can be easily adopted in problems on the knowledge bases in various sizes. The time cost is related to the sets of input and output. We can save time by controlling the threshold to diminish the size of these sets and finally save time.

\section{BATCH INTEGRATION}
\label{sec:batchIntegration}
In this section, we discuss batch integration implementation in detail. We first introduce the steps to construct the cluster set for batch integration in 6.1. How to resolve subset consisting unrelated schemas efficiently is provided in 6.2. 

\subsection{Flow of Batch Integration}
\label{sec:batchFlow}
According to Section~\ref{sec:overviewBatch}, the batch integration algorithm has four parts, initialization, ED Join, Semantic Join and resolve.

In the initialization step, we add all the attributes into set $U$, since they should be the center attributes and the join operation should be performed around them. Then we perform self-ED-Join on $U$ to eliminate literally similar attributes in batch. In this way, literally similar attributes are put into the same cluster sets. For example, we have attributes, ``word'', ``import'', ``name'', ``export'', ``work'', and ``nabe'', and the corresponding $U$ is structured as \{``word'', ``work''\}, \{``name'', ``nabe''\}, and \{``import'', ``export''\}.

Next, to aggregate semantic-related attributes, we perform Semantic Join on $U$. During join, the cluster sets sharing semantically similar attributes are merged into one. Thus, the cluster sets contain literally or semantically similar attributes. However, such sets do not meet the need of the problem proposed in Section~\ref{sec:problemDefinition}. Recall on the definition of the cluster set, the attributes in one set should be similar to each other. However, when merging different attributes to the cluster set, the added one is only similar to a part of attributes in the set, but have no relationship to others.

Thus, to make the schema integration accurate, we separate the cluster set to several smaller ones avoiding extra information, called resolve process. However, as we know, there may be a great number of attributes in a set. Thus, how to find the unrelated two attributes and resolve the set accurately and efficiently comes to a problem that impacts the integration quality.

\subsection{Resolving}
\label{sec:resolve}
After merging cluster sets by join algorithm, similar attributes have been clustered together. However, according to the discussion in Section~\ref{sec:batchFlow}, the cluster set is not a closure under the constraint predetermined sometimes. For example, we have following similar attributes under the threshold, ``house'' and ``home'', ``house'' and ``building'', ``home'' and ``family''. Therefore, we join them together. Apparently, it is ridiculous to consider ``building'' and ``family'' as related attributes according to manual judgement. They should be separated into different sets.

As mentioned in \cite{lee2000intelliclean}, such conflicts are caused by false positive, and such problem will magnify when the data size gets larger. It has decisive influence on the result. If non-related attributes $X_1$ and $X_2$ are put in one cluster set and regarded as similar attributes, only one of them, $X_1$ for example, should be kept in the global generated schema. However, this can lead to lost information of other attributes, i.e. $X_2$. Subgraph-based approaches could be applied to solve such problem as  \cite{wang2016efficient}. Although such solution can separate the cluster set by an efficient heuristic algorithm on large data sets, it is too complicated and need extra time and space cost. Here, we propose a simple solution which is efficient and functionally enough to solve the resolve problem. We use an example to illustrate the solution.

\begin{figure}
	\centering
	\includegraphics[width=200pt]{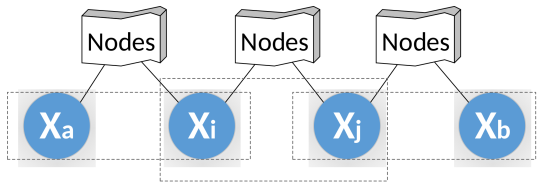}
	\caption{Example of Resolving}
	\label{fig:resolving}
\end{figure}

Figure~\ref{fig:resolving} shows some relation structure in cluster sets. Suppose that we perform join under a given threshold. The dashed line box contains attributes under the threshold. $dis(X_a, X_i)$, $dis(X_i, X_j)$ and $dis(X_j, X_b)$ are no more than the threshold, which means that they are similar attributes. Therefore, the attributes $X_a, X_i, X_j, X_b$ in the figure can be integrated into one cluster set. However, $X_a$ and $X_b$ have no relationship at all and $dis(X_a, X_b)$ is larger than the threshold. Such as the example above, ``building'' as $X_a$ and ``family'' as $X_b$ are non-related in fact, so we need to resolve $X_a$ and $X_b$ in the cluster set. Then a problem rises which cluster the attribute $X_i$, $X_j$ should be included.
Both partition \{\{$X_a, X_i, X_j$\},\{$X_b$\}\} and \{\{$X_b, X_i, X_j$\},\{$X_a$\}\} seem reasonable.
Also, is it possible to tolerant $X_a$ and $X_b$ in one set to reduce times of resolving when $dis(X_a, X_b)$ is not so large? Here, we give several resolving principles.

\begin{enumerate}
	\item To avoid relationship knowledge loss, such common attributes, $X_i$, $X_j$ in the example should be included in both the resolved sets.
	
	\item To decrease the amount of subsets, we choose not to keep the subsets as closures under given threshold $\gamma$. Instead, we define tolerance $\beta$ ($\beta>1$). During resolving the threshold in one cluster set in $\beta \gamma$. In this way with a large threshold, more attributes can be kept in the cluster set. $\beta$ can be defined from the pre-experiment.
	
	\item Other distance functions can be used in resolving, such as sum of squares of several path length under some constraint.
\end{enumerate}

\section{EXPERIMENTS}
\label{sec:experiment}
To verify the efficiency and effectiveness of the proposed approach, we conduct extensive experiments.

\subsection{Experimental Settings}
\label{sec:experimentalSettings}

\textbf{Environment.} The experiments are conducted on a computer of Windows 10 64 bit with an Intel Core i7 2.4GHz and memory of 8GB. All algorithms are implemented by VC++ of Visual Studio 2013 with single thread.

\textbf{Data Sets.} In order to test the accuracy and efficiency of our algorithm, we use real data from the Internet. For knowledge base, we choose \textit{Freebase} to provide knowledge linkage between concepts with special initialization mentioned following. Other knowledge base can be used as well.

For attributes of database tables to be integrated, we choose open data set from NYC OpenData\footnote{\url{https://data.cityofnewyork.us/}} and SF OpenData\footnote{\url{https://data.sfgov.org/}}.
These open data sets are sourced from real data. Such data cover various fields and test our algorithm comprehensively.

\textbf{Measures.} To compare the accuracy, recall and precision are commonly used to judge whether the results obtained from join schema integration are in high quality. $S_A$ is for the attributes which are found by our algorithm. $S_T$ is for the attributes related to the target determined by human. $S_A \cap S_T$ is the exactly related answer found by the integration algorithm. Also, $|S_A|$ means the total amount of such set of attributes, and the same to the others. Recall and precision are defined by the following formula.

\begin{eqnarray*}
	recall&=&\frac{|S_T \cap S_A|}{|S_T|}\nonumber\\
	precision&=&\frac{|S_T \cap S_A|}{|S_A|}\nonumber\\
\end{eqnarray*}

\textbf{Parameters.} Threshold $\varepsilon$ is a key parameter of join schema integration, which decides how many concepts are accessed during join. For two algorithms ED Join and Semantic Join, we set threshold $\varepsilon_t$ to describe how much misspelling can be tolerated while $\gamma$ determine the similarity degree. The values of these two thresholds are determined by the schema of attributes and the chosen knowledge base. To achieve a better performance and low cost, default value of these thresholds are $\varepsilon_t=1$ and $\gamma=3$.

\textbf{Initialization.}
In the experiments on real data, however, attributes in the database for experiment do not consist with the concepts in the knowledge base completely. Also, generally, the knowledge base cannot match each database table on the Internet. Therefore, we propose a way to make a link between them. Examples for attributes from database are shown as Table~\ref{tab:attributeExamples}.

\begin{table}
	\scriptsize
	\centering
	\caption{Attribute Examples}
	\begin{tabular}{cccc}
		\hline
		No. & Schema & Abbreviation & Rule\\
		\hline
		1 &	Name &  & a\\
		\hline
		2 &	ObjectID &  & c\\
		\hline
		3 &	FY 2011 Plan & FY(fiscal year) & b\\
		\hline
		4 &	Avg Speed & Avg(average) & b\\
		\hline
		5 &	Amount\_A &  & c\\
		\hline
		6 &	Report\_Num & Num(number) & b\\
		\hline
		7 &	Rpt\_Date & rpt(report) & b\\
		\hline
		8 &	Streetname &  & c\\
		\hline
		9 &	Jul-10 &  & d\\
		\hline
	\end{tabular}
	\label{tab:attributeExamples}
\end{table}

Following these examples and cases in real data sets, to make it consistent with concepts in the knowledge base, word correction is applied on the attribute names. To build a link between attributes and concepts, we proposed several rules for the transformation.

\begin{enumerate}
	\item Identity: easy words are no need to change, such as ``Name''. The attribute name is the same as concept name.
	
	\item Abbreviation: some attributes are given with an abbreviation rule in the schema description fields such as``FY''and ``Rpt'', or the common-used rules ``Num'' for ``number'' or ``Avg'' for ``average''. By using these rules, abbreviated attributes are expanded.
	
	\item Cutting words: For the words without obvious splitting signals, such as ``ObjectID'' and ``Streetname'', some dictionary are used as reference to separate them.
	
	\item Others: some words cannot match any certain inner part such as ``Jul-10'' which can be matched as ``Date''. Such other rules are offered manually based on common knowledge or a field-specified transformation rules.
\end{enumerate}

Following these rules, each attribute is transformed to several words. These transformation rules can also be generated in other ways such as \cite{arasu2009learning}. Then for matching, one of these words is to be chosen as the key of the expression and to be matched to the knowledge base concepts. We adopt tf-idf score~\cite{salton1983extended,sparck1972statistical}. The word with the highest tf-idf value in the expression is chosen as the keyword, and the final match rule is generated under such solution. Sometimes, keys from tf-idf are not so accurate to be on behalf of the attribute. Therefore, some skills like \textit{Textrank} in \cite{mihalcea2004textrank} or manual correction are required to achieve high accuracy.

%
%
%
%
%

\subsection{Case Study}
\label{sec:caseStudy}
In this section, we conduct case study to visually show how the join algorithms work. Since the concept size is too large, we only choose a small part of knowledge base for the convenience of case study. Also, for the ease of understanding, we do not include ED join based on misspelling of knowledge base and input in this section.

Firstly, we study a case of batch integration. Here, we have a schema set of pies and cookies, $W$ as \{``Pies'', ``Savory pies'', ``Sweet pies'', ``meat pie'', ``mince pie'', ``pie crust'', ``tart'', ``tartlet'', ``puff''\} and set the threshold as 2. After executing batch integration based on Semantic Join, we obtain the result set consisting of 102 attributes, most of which are food like ``Chocolate desserts'' and ``quiche'', food class like ``British pies'' and ``baked goods''. From this case, the attributes that integration by the algorithm is as excepted.

As for incremental integration, we have an empty existing set $\{U,S\}$, and all the attributes in $W$ (the same as batch case) are going to be integrated. When adding ``Pies'', ``Savory pies'', ``Sweet pies'', ``meat pie'', ``mince pie'', ``pie crust'' and ``tart'', there is nothing related in $U$, so the attributes are inserted into $U$ and their neighbors in the knowledge base are inserted into $S$. Then for ``tartlet'' as the neighbor of ``tart'' under threshold $\gamma$ in the knowledge base, it exists in $S$, and there is no need to be added. Then, to make the case fair enough, we try to integrate attributes in $W$ for another time. As expected, all attributes exist, and there is no need to expand $\{U,S\}$ furthermore.

\subsection{Accuracy}
\label{sec:accuracy}
To test how similar the results of the algorithm capturing and human idea is, we conduct experiments on data mentioned in Section~\ref{sec:experimentalSettings}. Manual integration results on the input attributes are used as the golden standard of this experiment. Although data sets have a large amount of attributes and concepts, for better analysis and manual judgment, we only use small data.

For batch integration, experimental results of different target attributes are shown in Table~\ref{tab:batchIntegrationAccuracy}. Recall and precision manifest that join schema integration works relatively well on different target words. The average recall and precision are 0.9266862 and 0.7431666, respectively. For better analysis on accuracy, the input set of batch integration is small to decrease manual judgement. When the size comes to 1, we can regard the problem as incremental integration. Hence, we do not conduct experiments with incremental integration here.

\begin{table}
	\scriptsize
	\centering
	\caption{Batch Integration Accuracy}
	\begin{tabular}{cccccc}
		\hline
		Word & $|S_A|$ & $|S_T|$ & $|S_T \cap S_A|$ & Recall & Precision\\
		\hline
		name &	76 & 61 & 57 & 0.934426 & 0.750000\\
		\hline
		year &	93 & 64 & 58 & 0.906250 & 0.617021\\
		\hline
		type &	73 & 58 & 53 & 0.913793 & 0.726027\\
		\hline
		number &	79 & 68 & 65 & 0.955882 & 0.822785\\
		\hline
		category &	12 & 13 & 15 & 0.923077 & 0.800000\\
		\hline
	\end{tabular}
	\label{tab:batchIntegrationAccuracy}
\end{table}

As for people, the similarity is decided by their background knowledge, and the relationship built based on the manual knowledge. As a simulation, absolutely, our proposed algorithm cannot make a perfect integration just as people because algorithms do not have similar knowledge as human. However, like the advantages of join algorithm shows, our algorithm can integrate schemas effectually. Some analysis is suggested as follows.

\begin{itemize}
	\item Recall Analysis\\
	From the result, we observe that the recall of integration is about 0.9. Recall indicates the coverage of result on human's idea. With the help of manual matching rules between attributes and knowledge concepts, integrated results considerably coincide with human, and recall is higher. To make it more objective, we conduct several experiments on different people and obtain a conclusive recall value in this result. However, this value can be humanly fluctuated both by background knowledge of human beings and the matching rules in Section~\ref{sec:experimentalSettings}.
	
	\item Precision Analysis\\
	Comparing to recall, precision is lower overall. After analyzing the words of database attributes and knowledge, concepts in the knowledge base cause the problem because the relationship in the knowledge base is not quantified. By getting knowledge from some like \textit{Freebase}, the relationship between concepts is described as an equal edge in the graph. Just as people think, both belonging to concept ``name'', we cannot regard ``first name'' and ``Peter'' at a same level of similarity. Obviously, ``first name'' is more related to concept ``name''. So under a certain threshold, some less-sense or even non-sense attributes are included as well, decreasing the precision value. To solve such problem, in some cases such as small subgraphs with only a few concepts, the precision is slightly affected when the threshold is low. More methods to imporve the knowledge base will be discussed in Section~\ref{sec:futureWork}.
\end{itemize}

\subsection{Efficiency}
\label{sec:efficiency}
For large-scale data, efficiency is extremely important. Therefore, we test the efficiency. From the algorithm, the running time is influenced by data size, target, threshold and the existence of cluster set. Hence, we show the experimental results and analyze their impact. In this section, we run the experiments on a real piece of knowledge from \textit{Freebase} containing 9,471,476 items.

\subsubsection{The Impact of Data Size}
\label{sec:dataSize}
To evaluate the impact of the size of data, we assign the size of input attributes at different values. The size of result attributes also change with it. The experimental result is shown in Table~\ref{tab:batchDataSize} for batch integration. It is easy to find running time is increasing with data growth. As for the result, we focus on the increasing rate of running time.

\begin{table}
	\centering
	\scriptsize
	\caption{Time Cost VS Data Size (batch)}
	\begin{tabular}{cccc}
		\hline
		\tabincell{c}{input\\ set No.} &	\tabincell{c}{input\\ size} &	\tabincell{c}{result\\ size} &	\tabincell{c}{running\\ time}\\
		\hline
		1 & 1 & 47 & 0.015\\
		\hline
		2 & 1 & 267 & 0.021\\
		\hline
		3 & 1 & 67476 & 0.672\\
		\hline
		4 & 5 & 12073 & 0.453\\
		\hline
		5 & 5 & 201529 & 5.625\\
		\hline
		6 & 10 & 106 & 0.084\\
		\hline
		7 & 10 & 19207 & 0.582\\
		\hline
		8 & 20 & 252163 & 51.13\\
		\hline
		9 & 20 & 84962 & 2.573\\
		\hline
		10 & 30 & 99243 & 12.39\\
		\hline
		11 & 30 & 177027 & 20.979\\
		\hline
		12 & 40 & 188034 & 30.185\\
		\hline
		13 & 40  & 376257 & 66.327\\
		\hline
		14 & 50 & 189247 & 18.009\\
		\hline
		15 & 50 & 204929 & 30.384\\
		\hline
	\end{tabular}
	\label{tab:batchDataSize}
\end{table}

For batch integration, we perform experiment on variation of input attribute size and result size. As observed from Table~\ref{tab:batchDataSize}, the result size has larger influence on running time than the input data size. For example, although the input set size is the same, the running time for input set 3 is larger than the input set 2 due to larger result size of the third. Even input set 5 has a larger running time than set 6 and 7, even though set 5 has a smaller input set. The reason is that integrating the attributes with many related concepts in the knowledge base means more time while accessing to the disk. Some input sets with less related knowledge cost less time. In conclusion, running time has no specific relationship with input size, but the result size makes much sense. The larger the result set is, the more information we can get form integration. By choosing a suitable threshold, we can limit the size of integrated attributes to save time while satisfying needed accuracy.

\subsubsection{The Impact of Target}
\label{sec:target}
\begin{figure}[!t]
	\centering
	\includegraphics[width=2.5in]{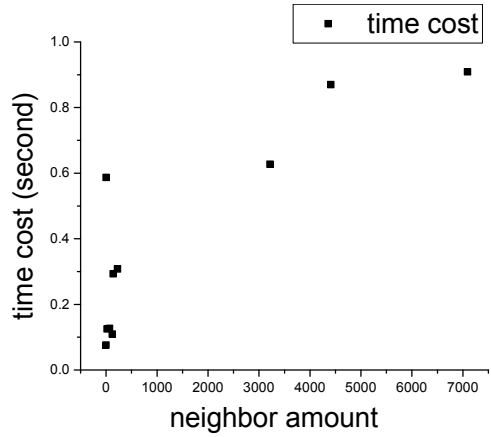}
	\caption{Time Cost VS Targets (incremental)}
	\label{fig:incrementalTarget}
\end{figure}

Since how much knowledge around a concept is unknown, we conduct experiments on concepts of different parts in the knowledge base to testify the impact of the target. Here, we select different attributes to conduct the experiment, both attributes with many neighbors and neighbor-less attributes. The running time is shown in Table~\ref{tab:batchTarget} for batch integration  and Figure~\ref{fig:incrementalTarget} for incremental integration. The running time differs quite a lot from these two types of attributes.

\begin{table}
	\centering
	\scriptsize
	\caption{Time Cost VS Targets (batch)}
	\begin{tabular}{cccc}
		\hline
		\tabincell{c}{input\\ set no.} &	\tabincell{c}{input\\ size} &	\tabincell{c}{result\\ size} &	\tabincell{c}{running\\ time}\\
		\hline
		1 & 1 & 75 & 0.031\\
		\hline
		2 & 1 & 3924 & 0.069\\
		\hline
		3 & 1 & 67476 & 0.672\\
		\hline
		4 & 5 & 377 & 0.108\\
		\hline
		5 & 5 & 12073 & 0.453\\
		\hline
		6 & 5 & 201529 & 5.625\\
		\hline
		7 & 10 & 106 & 0.084\\
		\hline
		8 & 10 & 19207 & 0.582\\
		\hline
		9 & 10 & 128659 & 4.463\\
		\hline
	\end{tabular}
	\label{tab:batchTarget}
\end{table}

\begin{table}
	\centering
	\scriptsize
	\caption{Time Cost VS Threshold (batch)}
	\begin{tabular}{cccc}
		\hline
		\tabincell{c}{input\\ size} &	threshold &	\tabincell{c}{result\\ size} &	\tabincell{c}{running\\ time}\\
		\hline
		1 & 2 & 66 & 0.024\\
		\hline
		1 & 3 & 213 & 0.129\\
		\hline
		1 & 4 & 69862 & 36.641\\
		\hline
		2 & 2 & 89 & 0.039\\
		\hline
		2 & 3 & 4145 & 0.304\\
		\hline
		2 & 4 & 108493 & 101.54\\
		\hline
		3 & 2 & 730 & 0.103\\
		\hline
		3 & 3 & 7086 & 3.481\\
		\hline
		3 & 4 & 111161 & 131.434\\
		\hline
	\end{tabular}
	\label{tab:batchThreshold}
\end{table}

From the table, we know that although input data and threshold are same, the running time can differ a lot. From the analysis in the experiment of data size, we know that result size has much influence. The experiment in this section verifies it as well. By analyzing the knowledge base, we can make sure why the variance of running time happens. In the knowledge base, the neighbor amount differs a lot among concepts. Some concepts belong to a small subgraph of neighbors, so there is a little knowledge to be dealt with during integration. On the other hand, if an attribute shares a lot of relationship with others, the problem can be very complex to enlarge the cluster set. For example, the input set 3 ``living people'' has much more neighbors (67476) that the input set 1 ``cancer'' (75) under threshold of 2 in Table~\ref{tab:batchTarget}. In conclusion, treating attributes with too much knowledge means much time to cost. To avoid large running time, when we foreknow the input attributes with many related attributes, we set a lower threshold, as discussed in Section~\ref{sec:threshold}.

\begin{figure}[!t]
	\centering
	\includegraphics[height=2.1in]{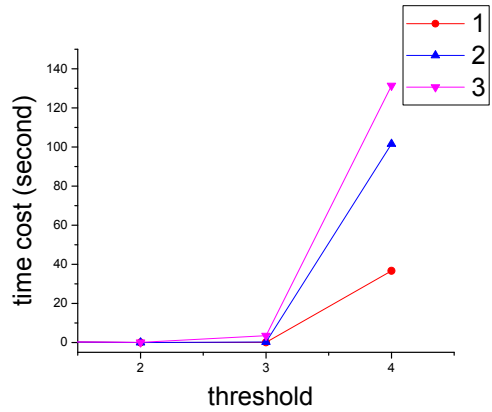}
	\caption{Time Cost VS Threshold (batch)}
	\label{fig:batchThreshold}
\end{figure}

For incremental integration, we test the running time when the target attribute is inserted. Based on the structure and cluster set mentioned in Definition~\ref{def:clusterSet}, if the added attributes are in the generated set $S$, it is unnecessary to insert and time can be saved. Here, we only consider the target with none relationship in $S$ and to be inserted to $U$. As observed in Figure~\ref{fig:incrementalTarget}, running time varies a lot when the neighbor amount is changed. Here, we select a part of the knowledge base and count the amount of 1-hop neighbor for each concept. The neighbor amount indicates that the concept comes from a dense subgraph or not. Figure~\ref{fig:incrementalTarget} shows that when the neighbor amount raises, the running time usually increases with it. Some outliers are caused by some 2-hop or more distant neighbor-rich concepts which are not presented by the neighbor size in Table~\ref{fig:incrementalTarget}.

\subsubsection{The Impact of Threshold}
\label{sec:threshold}
To test the impact of threshold, we set the threshold with different values. The minimum number of threshold is set to 2 to avoid too few integrated attributes. We test the trend of running time changing with the threshold. The experimental results for batch integration are shown as Table~\ref{tab:batchThreshold} and Figure~\ref{fig:batchThreshold}, respectively.

For batch integration, we conclude from the results that the value of threshold is very important. More integration attributes with larger threshold means more time or even being out of need. As the range of paths grows swiftly and the path set $P$ gets larger, the time of one loop of path expanding takes more time. We observe that the running time increases faster than linear time when the threshold goes beyond a certain value. 

Therefore, how the threshold value is set ahead of integration should be under deliberation, considering target, knowledge base, similar need, time need, etc. In practice, the threshold cannot be too high. Otherwise, the result attributes cannot be guaranteed similar as expected. As for the experiment with the threshold as 4, by scanning the result we know that many attributes are non-related to the input. For example, it is ridiculous to say that ``Dance festivals'' and ``mayor'' are related which is obtained by experiment in fact. However, when the threshold is set as 2 or 3, most of the result make sense. Therefore, the threshold should be set appropriately and reasonably.

As for incremental integration, the threshold also has influences when we perform ED Join and Semantic Join as talked above. From the aspect of join, we regard it as the batch integration with just one attribute. Therefore, when the threshold raises, the running time increases. However, when we execute the experiment, we observe that the set $S$ of cluster set is too large. Also actually, the size of $S$ is much large than $U$, and also much larger than need. Most of the attributes in $S$ is never accessed to compare for future attributes. Therefore, when inserting one attribute, we only choose to add some 1-hop neighbors with higher degree in the knowledge graph to $S$. Much time can be saved and it still works well as discussed in Section~\ref{sec:existing}.

\subsubsection{The Necessity of Cluster Set}
\label{sec:existing}
\begin{figure}
	\centering
	\includegraphics[height=2.1in]{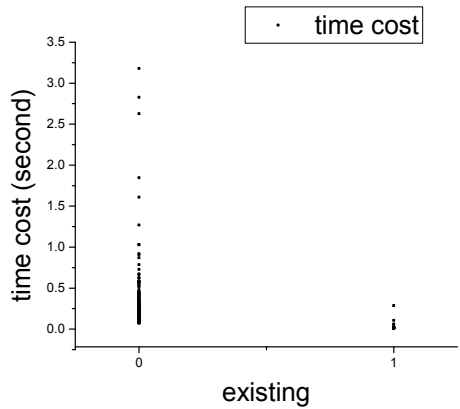}
	\caption{Time Cost VS Cluster Set Necessity (incremental)}
	\label{fig:increClusterSet}
\end{figure}

For incremental integration, for cluster set defined in Definition~\ref{def:clusterSet}, the generation of $S$ can save much time when the inserted attribute is related to existing ones in it. Here, we conduct some experiments to verify that the set $S$ really works for acceleration. The experimental results are shown in Figure~\ref{fig:increClusterSet}.

According to the results, we know that if one attribute exists in $S$, the running time for integrating such attribute keeps low. However, new attributes for the cluster set usually spends more time than those in $S$. Therefore, in this way, we can observe that the proposed structure, cluster set, can save time when the attributes appear for more than once.

\subsubsection{Conclusion}
\label{sec:efficiencyConclusion}
As is stated above, running time of join schema integration is complicated affected by data size, target, threshold and the existence of cluster set. To decrease the running time, it is necessary to balance these factors according to the requirement. For a certain problem, one good solution is to decrease the threshold as low as possible to limit the answer and save time.

%
\section{RELATED WORK}
\label{sec:relatedWork}
As a basic but crucial technique in database, schema integration has been discussed for many years. In old days, schema integration using similarity metric such as Jaccard similarity could not deal with semantic relation. Later, one marvelous work \cite{rahm2001survey} concludes many approaches to finish the work of schema mapping and integration. In this paper, the authors made classification for existing methods of schema integration and schema mapping, using techniques such as linguistic ways. For methods applied to schema integration, DIKE\cite{palopoli1998automatic} and ARTEMIS\cite{castano2001global} lead the ways. These two methods both computes the relationship between objects or attributes while our proposed algorithm use existing knowledge base. At most cases, relationship in knowledge base extracted from Web is in closer proximity to human's mind.

Recently, Microsoft has done some research \cite{He2015SEMA} on schema integration. In this paper, precision and recall of integration has a high value. Compared to our schema-level algorithm,  much instance information is used in their SEMA-JOIN. As the database tables have too many rows storing details, it is not possible to bring them all during the integration. For the efficiency, here, we choose to discard the instance information. What's more, there are quite a lot databases with less maintenance that have even no value for some attributes, integration in schema-level can be more widely used.



\section{CONCLUSIONS AND FUTURE WORK}
\label{sec:futureWork}

In this paper, we study a novel problem of schema integration on big data. To process this problem, we propose batch and incremental integration algorithms for different scenarios. The former is suitable for a set of attributes needed to be integrated, and the latter is used to insert information of newly adding attributes to the existing integrated cluster set. For effectiveness issues, we involve both semantics and syntactic similarity during integration. The semantics similarity is computed according to the knowledge based, and the syntactic similarity is based on the edit distance. For efficiency issues, we propose ED Join and Semantic Join algorithms. Experimental results show that our approaches could integrate schema efficiently and effectively.
%
%

Considering that current knowledge base actually cannot provide all needed information, our future work is to develop novel transformation rules discovery algorithms and weight determination algorithms for the knowledge base to achieve high accuracy for integration.

\paragraph*{Acknowledgement}

\bibliographystyle{plain}
\bibliography{SchemaIntegrationforBigData}
\end{spacing}{1.0}
\end{document}